\documentclass[12pt,onecolumn]{IEEEtran}
\usepackage{graphicx}
\usepackage{subfigure}
\usepackage{epsfig}
\usepackage{amssymb}
\usepackage{amsmath}
\usepackage{amsfonts}

\newcommand{\vp}{\mbox{${\bf p}$}}
\newcommand{\vd}{\mbox{${\bf d}$}}

\newcommand{\vx}{\mbox{${\bf x}$}}

\newcommand{\vt}{\mbox{${\bf t}$}}

\newcommand{\vs}{\mbox{${\bf s}$}}

\newcommand{\vr}{\mbox{${\bf r}$}}

\newcommand{\vn}{\mbox{${\bf n}$}}

\newcommand{\mB}{\hbox{{\bf B}}}

\newcommand{\mC}{\hbox{{\bf C}}}

\newcommand{\mH}{\hbox{{\bf H}}}

\newcommand{\mI}{\hbox{{\bf I}}}

\newcommand{\mR}{\mbox{{$\bf R$}}}

\newcommand{\ga}{\alpha}
\newcommand{\gb}{\beta}

\newcommand{\gz}{\zeta}

\newcommand{\gr}{\rho}
\newcommand{\gs}{\sigma}

\newcommand{\gD}{\Delta}

\def\bm#1{\mbox{\boldmath $#1$}}
\newcommand{\vga}{\mbox{$\bm \alpha$}}

\newcommand{\vgr}{\mbox{$\bm \rho$}}

\newcommand{\mgL}{\mbox{$\bm \Lambda$}}

\newcommand{\cR}{{\cal R}}

\newcommand{\SNR}{\ensuremath{\hbox{SNR}}}

\newcommand{\diag}{\ensuremath{\hbox{diag}}}

\newtheorem{theorem}{Theorem}[section]
\newtheorem{lemma}[theorem]{Lemma}

\newtheorem{prop}{Proposition}[section]
\newtheorem{claim}{Claim}[section]

\newtheorem{definition}{Definition}[section]
\newtheorem{question}{Question}[section]
\newtheorem{coro}{Corollary}[section]

\newcommand{\beq}{\begin{equation}}
\newcommand{\eeq}{\end{equation}}
\newcommand{\bea}{\begin{array}}
\newcommand{\ena}{\end{array}}
\newcommand{\bds}{\begin {itemize}}
\newcommand{\eds}{\end {itemize}}
\newcommand{\bdf}{\begin{definition}}
\newcommand{\blm}{\begin{lemma}}
\newcommand{\edf}{\end{definition}}
\newcommand{\elm}{\end{lemma}}
\newcommand{\bthm}{\begin{theorem}}
\newcommand{\ethm}{\end{theorem}}
\newcommand{\bprp}{\begin{prop}}
\newcommand{\eprp}{\end{prop}}
\newcommand{\bcl}{\begin{claim}}
\newcommand{\ecl}{\end{claim}}
\newcommand{\bcr}{\begin{coro}}
\newcommand{\ecr}{\end{coro}}
\newcommand{\bquest}{\begin{question}}
\newcommand{\equest}{\end{question}}

\newcommand{\rarrow}{{\rightarrow}}

\renewcommand{\baselinestretch}{1.5}
\begin{document}
\title{Cooperative game theory and the Gaussian interference channel}
\author{Amir Leshem ({\em Senior member}) and Ephraim Zehavi \thanks{
School of Engineering, Bar-Ilan University, Ramat-Gan,
52900, Israel. Part of this work has been presented at ISIT 2006 \cite{leshem2006}. 
This work was supported by Intel Corporation. e-mail: leshema@eng.biu.ac.il .}
({\em Senior member})}
\date{\today}
\maketitle

\begin{abstract}
In this paper we discuss the use of cooperative game theory for analyzing
interference channels. We extend our previous work, to games with $N$ players
as well as frequency selective channels and joint TDM/FDM strategies.

We show that the Nash bargaining solution can be computed using convex
optimization techniques. We also show that the same results are applicable to
interference channels where only statistical knowledge of the channel is
available. Moreover, for the special case of two players $2\times K$ frequency
selective channel (with K frequency bins) we provide an $O(K \log_2 K)$
complexity algorithm for computing the Nash bargaining solution under mask
constraint and using joint FDM/TDM strategies. Simulation results are also
provided.

\end{abstract}
\section{Introduction}
Computing the capacity region of the interference channel is an open problem in information
theory \cite{cover}.
A good overview of the results until 1985 is given by van der Meulen  \cite{Meulen94}
and the  references therein. The capacity region of general interference case
is not known yet. However, in the last forty five years of research some
progress has been made. Ahslswede \cite{ahlswede73}, derived a general
formula for the capacity region of a discrete memoryless  Interference
Channel (IC) using a limiting expression which is computationally infeasible.
Cheng, and  Verdu \cite{cheng93} proved that the
limiting expression cannot be written in general by a single-letter formula
and  the restriction to Gaussian inputs provides only an inner bound to the
capacity region of the IC. The best known achievable region for the general
interference channel is due to Han and Kobayashi  \cite{han81}. However the
computation of the Han and Kobayashi formula for a general discrete memoryless
channel is in general too complex. Sason \cite{sason2004} describes certain improvement over the Han Kobayashi rate region in certain cases.
A 2x2 Gaussian interference channel
in standard form (after suitable normalization) is given by:
\beq
\label{standard_IC}
\vx=\mH \vs +\vn, \quad
\mH=\left[
\bea{cc}
1 & \ga \\
\gb & 1
\ena
\right]
\eeq
where, $\vs=[s_1,s_2]^T$, and $\vx=[x_1,x_2]^T$ are sampled values of the input and
output signals, respectively. The noise vector $\vn$ represents the
 additive Gaussian noises with zero mean and unit variance. The powers of the
input signals are constrained to be less than $P_1,P_2$ respectively. The off-diagonal elements of $\mH$, $\ga,\gb$ represent the degree of interference
present. The capacity region of the Gaussain interference channel with very
strong interference (i.e., $\ga \ge 1+P_1$, $\gb \ge 1+P_2$ )
was found by Carleial
given by
\beq
\label{VSI_RR}
R_i \le \log_2(1+P_i), \ \ i=1,2.
\eeq
This surprising result shows that very strong interference dose not reduce the capacity.
A Gaussian interference channel is said to
have strong interference if  $\min\{\ga,\gb\}>1$. Sato
\cite{sato81} derived an achievable capacity region (inner bound) of Gaussian
interference channel as intersection of two multiple access gaussian capacity
regions embedded in the interference channel. The achievable region is the
intersection of the rate pair  of the rectangular region of the very strong
interference (\ref{VSI_RR}) and the region
\beq
R_1+R_2 \le \log_2\left(\min\left\{1+P_1+\ga P_2,1+P_2+\gb P_1\right\} \right).
\eeq
A recent progress for the case of Gaussian interference is described by Sason  \cite{sason2004}.
Sason derived an achievable rate region based on a modified time- (or frequency-) division
multiplexing approach which was originated by Sato for the degraded Gaussian IC. The achievable
rate region includes the rate region which is achieved by time/frequency division multiplexing
(TDM/ FDM), and it also includes the rate region which is obtained by time sharing between the
two rate pairs where one of the transmitters sends its data reliably at the maximal possible
rate (i.e., the maximum rate it can achieve in the absence of interference), and the other
transmitter decreases its data rate to the point where both receivers can reliably decode their messages.

While the two users fixed channel interference channel is a well studied problem, much less is known in the
frequency selective case. An  $N\times N$ frequency selective Gaussian
interference channel is given by:
\beq
\label{standard_IC_matrix}
\bea{c}
\vx_k=\mH_k \vs_k +\vn_k
\qquad k=1,...,K \\
\mH_k=\left[
\begin{array}{ccc}
             h_{11}(k) & \hdots &  h_{1N}(k)\\
             \vdots & \ddots & \vdots \\
             h_{N1}(k) & \hdots &  h_{NN}(k)\\
\end{array}
\right].
\ena
 \end{equation}
where, $\vs_k$, and $\vx_k$ are sampled values of the input and output signal vectors at
frequency $k$, respectively. The noise vector $\vn_k$ represents the
 additive Gaussian noises with zero mean and unit variance. The power spectral density (PSD)
of the input signals are constrained to be less than $p_1(k),p_2(k)$
respectively. The off-diagonal elements of $\mH_k$, represent the
degree of interference present at frequency $k$. The main difference between interference channel
and a multiple access channel (MAC) is that in the interference channel, each component
of $s_k$ is coded independently, and each receiver has access to a single element of
$\vx_k$. Therefore iterative decoding schemes are much more limited, and typically
impractical.

One of the simplest ways to deal with interference channel is
through orthogonal signaling. Two extremely simple orthogonal schemes are using
FDM or TDM strategies. For frequency selective channels (also known
as ISI channels) we can combine both strategies by allowing time
varying allocation of the frequency bins to the different users. In
this paper we limit ourselves to joint FDM and TDM scheme where an
assignment of disjoint portions of the frequency band to the several
transmitters is made at each time instance. This technique is widely
used in practice because simple filtering can be used at the
receivers to eliminate interference. In this paper we will assume a
PSD mask limitation (peak power at each frequency) since this constraint is typically
enforced by regulators.

While information theoretical considerations allow all points in the rate region, we argue that
the interference channel is a conflict situation
between the interfering links \cite{leshem2006}. Each link is considered a player in a
general interference game.  As such it has been shown that non-cooperative
solutions such as the iterative water-filling, which leads to good solutions
for the multiple access channel (MAC) and the broadcast channel \cite{yu04}
can be highly suboptimal in interference  channel scenarios \cite{laufer2005},
\cite{laufer2005a}. To solve this problem
there are several possible approaches. One that has gained popularity in
recent years is through the use of competitive strategies in repeated games
 \cite{etkin2005}. Our approach is significantly
 different and is based on general bargaining theory originally developed by
Nash \cite{owen}. Our approach is also different than that of \cite{mathur2006}
where Nash bargaining solution for interference channels is studied under the assumption
of receiver cooperation. This translates the channel into a MAC, and is not relevant to
distributed receiver topologies. In our analysis of the interference channel we claim
that while all points on the boundary of the interference channel are achievable from
the  strict informational point of view, many of them will never be achieved
since one of the players will refuse to use coding strategies
leading to these points.
 The rate vectors of interest are only rate vectors that dominate component-wise
 the rates that each user can achieve, independently of the other users coding  strategy.
The best rate pairs that can be achieved independently of the other
users strategies form a Nash equilibrium \cite{owen}. This implies
that not all the rates are indeed achievable from game theoretic prespective.
Hence we define the game theoretic rate region. \bdf Let $\cR$ be an
achievable information theoretic rate region. The game theoretic
rate region ${\cR}^G$ is given by
 \beq
 \cR^G=\left \{
 (R_1,...,R_N)\in \cR: R^c_i \le R_i, \ \ \hbox{for all \ } i=1,...,N \right\}
 \eeq
 where $R^c_i$ is the rate achievable by user $i$ in a non-cooperative
interference game \cite{laufer2005a}. \edf
 To see what are the pair rates that can be achieved by negotiation and cooperation
 of the users we resort to a well known solution termed the Nash bargaining
 solution. In his seminal papers, Nash proposed four axioms required that any
 solution to the bargaining problem should satisfy. He then proved that there
 exists a  unique solution satisfying these axioms. We will analyze the
 application of Nash bargaining solution (NBS) to the interference game, and
 show that there exists a unique point on the boundary of the capacity region
 which is the solution to the bargaining problem as posed by Nash.

 The fact that the Nash  solution can be computed independently by users, using
 only channel state information, provides a good method for managing
 multi-user ad-hoc networks operating in an unregulated environment.

Application of Nash bargaining to  OFDMA has been proposed by
\cite{han2005}. However in that paper the solution was used only as
a measure of fairness.  Therefore, $R^c_i$ was not taken as the Nash
equilibrium for the competitive game, but an arbitrary $R^{\min}_i$.
This can result in non-feasible problem, and the proposed algorithm
might be unstable. The algorithm in \cite{han2005} is suboptimal even
in the two users case, and according to the authors can lead to an
unstable situation, where the Nash bargaining solution is not
achieved even when it exists. In contrast, in this paper we show that the NBS 
for the $N$ palyers game can be computed using convex optimization techniques. 
We also provide detailed analysis of
the two users case and provide an $O(K \log_2 K)$ complexity algorithm
which provably achieves the joint FDM/TDM Nash bargaining solution.
Our analysis provides ensured convergence for higher
number of users and bounds the loss in applying OFDMA compared to
joint FDM/TDM strategies. In the two users case we
can show that the Nash bargaining solution requires TDM over no more
than a single tone, so we can achieve a very good approximation to
the optimal FDM based Nash bargaining solution. We also provide similar
analysis for higher number of users, showing that for the Nash bargaining solution with
$N$ players, over a frequency selective channel with $K$ frequency bins, only
${{N} \choose {2}}$ frequency bins has to be shared by TDM, while all other frequencies are
allocated to a single user. When
${{N} \choose {2}} <<K$, this provides a near optimal solution to the game using FDM
strategies, as well.

The structure of the paper is as follows: In section \ref{game_theory}
we discuss competitive and cooperative solutions to interference games and provides an
overview of the Nash bargaining theory. In section \ref{flat_fading} we discuss the
existence of the NBS for $N$ players FDM cooperative game over slow, flat fading channels.
In section \ref{freq_selective} we discuss the Nash bargaining over general frequency selective
interference channel, with mask constraint. We show that computing the NBS under mask
constraint and joint FDM/TDM strategies can be posed as a convex optimization problem. This
shows that even for large number of palyers, computing the solution with many tones is
feasible. We also show that in this case the $N$ users will share only few frequencies,
dividing all the others. In section \ref{two_players_section} we specialize to the two players case, but
with frequency selective channels. We provide an algorithm for computing the NBS in
complexity $O(K \log_2(K)$.
Finally, we demonstrate in simulations the gains compared to to the competitive solution
both in the flat fading and the frequency selective cases. We end up with some conclusions.
\section{Nash equilibrium vs. Nash Bargaining solution}
\label{game_theory}
In this section we describe two solution concepts for $N$ players games. The
first notion is that of Nash equilibrium. The second is the Nash bargaining
solution (NBS).
In order to simplify the notation we specifically concentrate on the Gaussian
interference game.

 \subsection{The Gaussian interference game}
 \label{sec:GI_game} In this section we define the Gaussian
 interference game, and provide some simplifications for dealing with
 discrete frequencies. For a general background on non-cooperative
 games we refer the reader to \cite{owen}.
 The Gaussian interference game was defined in
 \cite{yu2002}. In this paper we use the discrete approximation
 game. Let $f_0 < \cdots <f_K$ be an increasing sequence of
 frequencies. Let $I_k$ be the closed interval be given by
 $I_k=[f_{k-1},f_k]$. We now define the approximate Gaussian
 interference game denoted by $GI_{\{I_1, \ldots, I_K\}}$.

 Let the players $1,\ldots,N$ operate over $K$ parallel channels. Assume
 that the $N$ channels have transfer functions $h_{ij}(k)$.
 Assume that user $i$'th is allowed to transmit a total power of
 $P_i$. Each player can transmit a power vector $\vp_i=\left(
 p_i(1),\ldots,p_i(K) \right)  \in [0,P_i]^K$ such that $p_i(k)$ is
 the power transmitted in the interval $I_k$. Therefore we have
 $\sum_{k=1}^K p_i(k)=P_i$. The equality follows from the fact that in
 non-cooperative scenario all users will use the maximal power they
 can use. This implies that the set of power distributions for all
 users is a closed convex subset of the cube $\prod_{i=1}^N
 [0,P_i]^K$ given by:
 \beq \label{eq_strategies} \mB=\prod_{i=1}^N
 \mB_i
 \eeq where $\mB_i$ is the set of admissible power
 distributions for player $i$ given by:
 \beq
 \mB_i=[0,P_i]^K\cap
 \left\{\left(p(1),\ldots,p(K)\right): \sum_{k=1}^K p(k)=P_i \right\}.
 \eeq
 Each player chooses a PSD $\vp_i=\left<p_i(k): 1\le k \le N
 \right > \in \mB_i$. Let the payoff for user $i$ be given by:
 \beq
  \label{eq_capacity}
 \begin{array}{l}
 C^i\left(\vp_1,\ldots,\vp_N\right)=
 \sum_{k=1}^{K}\log_2\left(1+\frac{|h_i(k)|^2p_i(k)}{\sum
 |h_{ij}(k)|^2 p_j(k)+\gs^2_i(k)}\right)
 \end{array}
 \eeq
 where $C^i$ is the capacity available to player $i$ given power
 distributions $\vp_1,\ldots,\vp_N$, channel responses $h_i(f)$,
 crosstalk coupling functions $h_{ij}(k)$ and $\gs^2_i(k)>0$ is external
 noise present at the $i$'th receiver at frequency $k$. In
 cases where $\gs^2_i(k)=0$ capacities might become infinite using FDM
 strategies, however this is non-physical situation due to the
 receiver noise that is always present, even if small. Each $C^i$ is
 continuous on all variables.

 \begin{definition}
 The Gaussian Interference game $GI_{\{I_1,\ldots,I_k\}}=\left\{\mC,\mB\right\}$ is the N
 players non-cooperative game with payoff vector
 $\mC=\left(C^1,\ldots,C^N \right)$ where $C^i$ are defined in
 (\ref{eq_capacity}) and $\mB$ is the strategy set defined by (\ref{eq_strategies}).
 \end{definition}
 The interference game is a special case of convex non-cooperative N-persons
 game.
\subsection{Nash equilibrium in non-cooperative games}
An important notion in game theory is that of a Nash
 equilibrium.
 \bdf
 An $N$-tuple of strategies $\left<\vp_1,\ldots,\vp_N\right>$ for
 players $1,\ldots,N$ respectively
 is called a Nash equilibrium iff for all $n$ and for all $\vp$ ($\vp$ a
 strategy for player $n$)
 \[
 C^n\left(\vp_1,...,\vp_{n-1},\vp,\vp_{n+1},\ldots,\vp_N \right)<
 C^n\left(\vp_1,...,\vp_{N} \right)
 \]
 i.e., given that all other players $i \neq n$ use strategies $\vp_i$, player
 $n$ best response is $\vp_n$.
 \edf
 The proof of existence of Nash equilibrium in the general interference
 game follows from an easy adaptation of the proof of the this result
 for convex games \cite{leshem2006}.
 A much harder problem is the uniqueness of Nash equilibrium points in
 the water-filling game. This is very important to the stability of the
 waterfilling strategies. A first result in this direction has been
 given in \cite{yu2000}, \cite{chung2002}. 
 A more general analysis of the convergence
 has been given in  \cite{luo2005}.

 \subsection{Nash bargaining solution for the interference game}
Nash equilibria are inevitable whenever a non-cooperative zero sum
 game is played. However they can lead to substantial loss to all players,
compared to a  cooperative strategy in the non-zero sum case, where players
can cooperate. Such a situation is called the prisoner's dilemma.
The main issue in this case is how to achieve the cooperation in a stable
manner and what rates can be achieved through cooperation.

In this section we present the Nash bargaining solution \cite{owen}.
The underlying structure for a Nash bargaining in an $N$ players
game is a set of outcomes of the bargaining process $S$ which is
compact and convex. $S$ can be considered as a set of possible joint
strategies or states, a designated disagreement outcome $d$ (which
represents the agreement to disagree and solve the problem
competitively) and a multiuser utility function $ U:S \cup \{d\}
\rarrow {\bm R}^N. $ The Nash bargaining is a function $F$ which
assigns to each pair $\left(S \cup \{d\},U \right)$ as above an
element of $S \cup \{d\}$. Furthermore, the Nash solution is unique.
In order to obtain the solution, Nash
 assumed four axioms:
\bds
\item[]{\em Linearity}. This means that if we perform  the same
linear transformation on the utilities of all players than the solution is
transformed accordingly.
\item[]{\em Independence of irrelevant alternatives}. This axiom states that
if the bargaining solution of a large game $T \cup \{d\}$ is obtained in a
small
set $S$. Then the bargaining solution assigns the same solution to the smaller game, i.e., The irrelevant alternatives in $T \backslash S$ do not affect the outcome of the bargaining.
\item[]{\em Symmetry}. If two players are identical than renaming them will not
change the outcome and both will get the same utility.
\item[]{\em Pareto optimality}. If $s$ is the outcome of the bargaining then no other
state $t$ exists such that $U(s)<U(t)$ (coordinate wise).
\eds
A good discussion of these axioms can be found in \cite{owen}.
Nash proved that there exists a unique solution to the bargaining problem
satisfying these 4 axioms. The solution is obtained by maximizing
\beq
s= \arg \max_{s\in S \cup \{d\}} \prod_{n=1}^N \left(U_n(s)-U_n(d) \right).
\eeq
Typically one assumes that there exist at least one feasible
 $s \in S$ such that $U(d)<U(s)$ coordinatewise, but otherwise we can assume that the
bargaining solution is $d$.

We also define the Nash function $F(s):S \cup \{d\} \rarrow {\boldmath R}$
\beq
\label{nash_function}
F(s)=\prod_{n=1}^N \left(U_n(s)-U_n(d) \right).
\eeq
The Nash bargaining solution is obtained by maximizing the Nash function over all possible
states. Since the set of possible outcomes $U\left( S \cup \{d\}  \right)$ is convex $F(s)$
has a unique maximum on the boundary of $U\left( S \cup \{d\}  \right)$.

Whenever the disagreement situation can be decided by a competitive game, it is reasonable
to assume that the disagreement state is given by a Nash equilibrium of the relevant
competitive game. When the utility for user $n$ is given by the rate $R_n$, and
$U_n(\vd)$ is the competitive  Nash equilibrium, it is obtained  by
iterative waterfilling for general ISI channels. For the case of
mask constraints the competitive solution is simply given by all
users using the maximal PSD at all tones.
%
%
\section{Nash bargaining solution for the flat fading $N$ players interference game }
\label{flat_fading}
 In this section we provide conditions for the existence of the Nash bargaining solution
 (NBS) for the $N\times N$ flat frequency interference game. In general, the rate region for the
 interference channel is unknown. However, by a simple time sharing argument we know that
 the rate region is always a convex set $\cR$,
 i.e.
\beq \cR=\left\{\textbf{r}: \textbf{r}=\left(R_1,R_2,...,R_N
\right)\texttt{is in the rate region }\right\}.\eeq
is a convex set. Typically we will use the utility defined by the rate, i.e., for every rate vector $\vr=(R_1,...,R_N)^T$
we have $U_n(\vr)=R_n$. Later we will show how the results can be generalized to 
other utility functions such as $U^L_n(\vt)=\log\left(R_n\right)$

For some specific operational strategies one can define an
achievable rate region explicitly. This allows for explicit determination of the strategies 
leading to  the NBS. One such example is the use of FDM or TDM strategies in the interference 
channel. In the sequel we analyze the $N$ players interference game, with FDM
or TDM strategies. We provide conditions under which the bargaining
solution exists, i.e., FDM strategies provide improvement over the competitive solution. This
extends the work of \cite{laufer2005} where we characterized when
does FDM solution outperforms the competitive IWF solution for
symmetric 2x2 interference game. We have shown there that indeed in
certain conditions the competitive game is subject to the prisoner's
dilemma where the competitive solution is suboptimal for both
players. Let the utility of player $n$ is given by $U_n=R_n$.
 The received signal vector $\vx$ is given by
 \beq
 \vx=\mH \vs + \vn
 \eeq
where $\vx=[x_1,...,x_N]^T$ is the received signal, and
$\mH=\{h_{ij}\}, 0\leq i,j\leq N$, is the interference coupling matrix,
$\vs=[s_1,s_2,...,s_N]^T$ is the vector of transmitted signals.
We will assume that for all $i,j$ $|h_{ij}|<1$. Moreover, we will assume
that the matrix $\mH$ is invertible. This assumption is reasonable since typical 
wireless communication channels are random, and the probability of obtaining a singular 
channel is 0.
Note that in our case
both transmission and reception are performed independently, and the
vector formulation is used for notational simplicity. First observe:
\blm
The competitive strategies in the Gaussian interference game are given by flat power
allocation. The resulting rates are:
 \beq
 \label{flat_power}
 R^c_{n}=\frac{W}{2}
 \log_2 \left(1+\frac{|h_{nn}|^2 P_{n}}{WN_0/2+\sum_{j=1,j\neq n}^N|h_{nj}|^2 P_{ij}}\right)
 \eeq
 \elm
 {\bf Proof:}
 To see that the flat power allocations form a Nash equilibrium for a flat channel,
 we first note that when all players $j \neq n$ use flat power spectrum, the total
 interference plus noise spectrum is also flat. Hence waterfilling by player $n$ against
 flat power allocation results in flat power spectrum. This implies that the flat power
 spectrum is indeed a Nash equilibrium point.
 To obtain the uniqueness, assume that the total power limit of the users is given by
 $\vp=[P_1,...,P_N]^T$ and that the spectrum is divided into $K$ identical bands.
 Assume that user $n$ strategy at the equilibrium is given by
 $\vgr=[\gr_n(1),...,\gr_n(K)]^T$. We note that the mutual
 waterfilling equations can be written for all $k \neq k'$
 \beq
 \mH \mgL_k \vp + N_0\mI= \mH \mgL_{k'} \vp+N_0 \mI
 \eeq
 where $\mgL_k=\diag\{\gr_1(k),\ldots,\gr_N(k)\}$.
 By our assumption $\mH$ is invertible and $\mgL_k$ is diagonal for each $k$ so we must have
 for all $n,k$, $\gr_n(k)=\gr_n(1)$, obtaining the uniqueness.
Finally we note that when interference is very strong there are other Nash equilibrium points
on the boundary of the strategy space, where not all frequencies are used by all users.

 To simplify the expression for the competitive rates we divide the expression inside the 
 $\log$ in (\ref{flat_power}) by the noise power $WN_0/2$ obtaining:
 \beq
\label{competitive}
 R^c_n=\frac{W}{2}\log_2 \left(1+\frac{\SNR_n}{1+\sum_{j\neq n}^N\ga_{nj}\SNR_j} \right)
 \eeq
 where
 $
 \SNR_j=\frac{|h_{jj}|^2 P_j}{WN_0/2},
 \ga_{nj}=\frac{|h_{nj}|^2}{|h_{jj}|^2}.
 $
 Since the rates $R^c_n$ are achieved by competitive strategy, player
$n$ would not cooperate unless he will obtain a rate higher than
$R^c_n$. Therefore, the game theoretic rate region is defined by set
of rates higher that $R_n^c$ of equation (\ref{competitive}).

We are interested in FDM cooperative strategies. 
A strategy is a vector $[\gr_1,...,\gr_N]^T$
such that $\sum_{n=1}^N \gr_n \le 1$. We assume that
player $n$ uses a fraction $\gr_n$ $\left(0 \le \gr_n \le 1\right)$ of the band 
(or equivalently uses the channel for a fraction $\gr_n$ of the time in the TDM case). 
The rate obtained by the $n^{th}$ player is given by
 \beq
 \label{def_Rrho}
 \bea{l}
 R_n(\vgr)=R_n(\gr_n)=\frac{\gr_n W}{2} \log_2 \left(1+\frac{\SNR_n}{\gr_n} \right).
 \ena
 \eeq
 First we note that the FDM rate region $\cR_{FDM}=\left\{(R_1,...,R_N)| R_n=R_n(\gr_n) \right\}$ 
 is indeed convex. The Pareto optimal points must satisfy $\sum_{n=1}^N\gr_n=1$, since
 by dividing the unused part of the band between users, all of them increase their utility.
  Also note that by strict monotonicity of $R_n(\gr)$ as a function of $\gr$ 
 each pareto optimal point is on the boundary of $\cR_{FDM}$. It is achieved by a single 
 strategy vector $\vgr$.
 Player $n$ benefits from FDM cooperation as long as
 \beq
 \label{ineqa}
 R^c_n < R_n(\gr_n).
 \eeq
 The Nash function is given by
 \beq
 F(\vgr)=\prod_{n=1}^N\left(R_n(\gr_n)-R_n^c \right).
 \eeq
To better understand the gain in FDM strategies we define a function $f(x,y)$ that is 
fundamental to the analysis.
 \bdf
 For each $0<x,y$ let $f(x,y)$ be defined by
\beq
\label{eq:def_f}
 f(x,y)=\min\left\{\gr:  \left(1+\frac{x}{\gr} \right)^\gr=1+\frac{x}{1+y}
\right\}.
\eeq
\edf
\bcl
\bds
\item[1.] $f(x,y)$ is a well defined function for $x,y \in {{\bm R}^+}$. 
\item[2.] For all $x,y \in {{\bm R}^+}$, $0<f(x,y)<1$.
\item[3.] $f(x,y)$ is monotonically decreasing in y.
\eds
\ecl
{\bf Proof:} Let $g(x,y,\gr)$ be defined by:
 \[
 g(x,y,\gr)=\left(1+\frac{x}{\gr} \right)^\gr-1-\frac{x}{1+y}
 \]
 For every $x,y$, $g(x,y,\gr)$ is a continuous and monotonic function in $\gr$.
 Furthermore, for any $0 < x,y$, $g(x,y,1)>0$, and
$\lim_{\gr \rarrow 0} g(x,y,\gr)<0$.
Hence, there is a unique solution to (\ref{eq:def_f}). Furthermore, the value 
of $f(x,y)$ is
strictly between $0,1$. Finally $f(x,y)$ is monotonically decreasing in $y$ since 
$g(x,y,\gr)$ is increasing in $y$, so if we increase $y$ we need to decreas $\gr$ to 
maintain a fixed value.

Using the function $f(x,y)$ we can completely characterize the cases where $NBS$ is 
preferable to the Nash equilibrium.
\bthm 
\label{theorem2}
Nash bargaining solution exists if and only if the following inequality holds
\beq \label{theorem2e}
\sum_{n=1}^N
f\left(\SNR_n,\sum_{j\neq n}\ga_{nj}\SNR_j\right) \le 1.
\eeq
 \ethm
 Proof: In one direction, assume that a Nash bargaining solution exists.
 The next two conditions must hold
 \bds
\item[1.] There is a partition of the band between the players such that player $n$
gets a fraction $\gr_n>0$.
\item[2.] Each player gets by cooperation higher rate then the competitive rate,
 i.e, $ R_n(\gr_n) \geq R^c_n$.
 \eds
 Therefore, using equation (\ref{def_Rrho}) and inequality (\ref{ineqa})
  we obtain that equation (\ref{theorem2e}) must be satisfied.
 On the other direction by definition of $f$ player $n$ has at least the rate that it can
 get by competition if he can use a fraction  $\gr_n$,  of the bandwidth.
 Since  (\ref{theorem2e}) implies that $\sum_{n=1}^N\gr_n \le 1$, FDM is preferable to the
competitive solution for the utility function $U_n=R_n$. By the convexity of the FDM rate
region the Nash function has a unique maximum that is Pareto optimal and outperforms the
competitive solution.

Interestingly, as long as the utility function $U_n(\vgr)$ depends only on $\gr_n$ and
$U_n(\gr)$ is monotonically increasing in $\gr$ the same conclusion holds. This implies that 
the NBS when the utility is $U^L_n(\gr)=\log\left(R_n(\gr_n)\right)$ there is a unique 
frequency division vector $\vgr$ that achieves the NBS. Furthermore the optimization problem, of 
computing the optimal $\vgr$ is still convex.

\begin{figure}
\begin{center}
\mbox{\psfig{figure=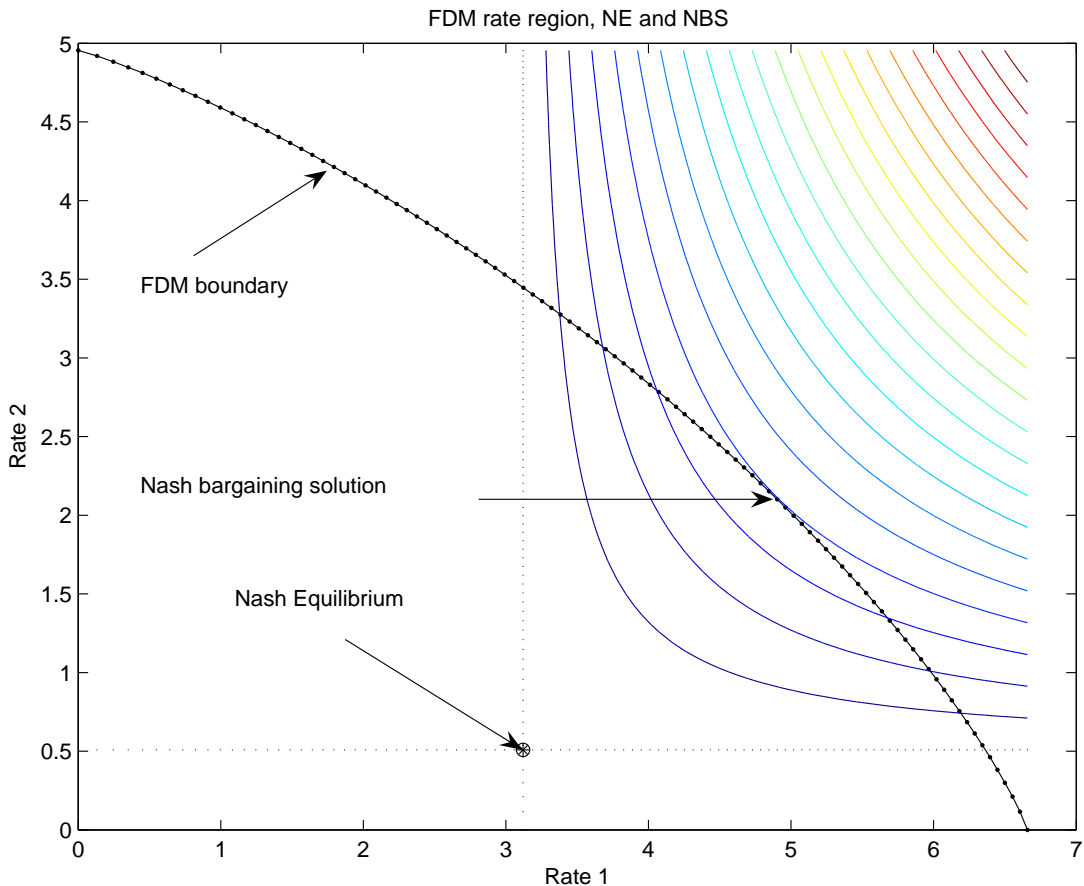,width=0.8\textwidth}}
\end{center}
    \caption{FDM rate region (thick line), Nash equilibrium $*$, Nash bargaining solution and the contours of $F(\gr)$. $\SNR_1=20$ dB, $\SNR_2=15$ dB,
and $\ga=0.4, \gb=0.7$ } \label{rate_region}
\end{figure}

We now examine the simple case of two players. Assume
that player I uses a fraction $\gr$ $\left(0 \le \gr \le 1\right)$ of the band and
user II uses a fraction $1-\gr$. The rates obtained by the two users
are given by
 \beq
 \label{def_Rrho}
 \bea{l}
 R_1(\gr)=\frac{\gr W}{2} \log_2 \left(1+\frac{\SNR_1}{\gr} \right) \\
 R_2(1-\gr)=\frac{(1-\gr) W}{2} \log_2 \left(1+\frac{\SNR_2}{1-\gr} \right)
 \ena
 \eeq
 The two users will benefit from FDM cooperation as long as
 \beq
 \bea{l}
 R^c_i \le R_i(\gr_i), \ \ \ i=1,2 \\
 \gr_1+\gr_2 \le 1
 \ena
 \eeq
Condition (\ref{theorem2e}) can now be simplified:
 \beq
\label{FDM_better}
 f(\SNR_1,\ga \SNR_2)+f(\SNR_2,\gb \SNR_1) \le 1,
 \eeq
 where
 \[
 \bea{lclcl}
 \SNR_i=\frac{|h_{ii}|^2 P_i}{WN_0/2},&\ \ &
 \ga=\frac{|h_{12}|^2}{|h_{22}|^2},& \ \  &
 \gb=\frac{|h_{21}|^2}{|h_{11}|^2}
 \ena.
 \]
The NBS is given by solving the problem
 \beq
 \gr_{NBS}=\arg \max_\gr F(\gr)
 \eeq
 where the Nash function is now given by:
\beq \label{def_F} F(\gr)=\left(R_1(\gr)-R^c_1 \right)
\left(R_2(1-\gr)-R^c_2 \right) \eeq and $R_i(\gr)$ are defined by
(\ref{def_Rrho}).
 A special case can now be derived:
 \bcl
 Assume that $\SNR_1 \ge \frac{1}{2} \left( \ga^2 \gb^4\right)^{-1/3}$ and
 $\SNR_2 \ge \frac{1}{2} \left( \gb^2 \ga^4\right)^{-1/3}$. Then there is a Nash
 bargaining solution that is better than the competitive solution. When the channel is
 symmetric ($\ga=\gb)$ the solution exists as long as $SNR \ge \frac{1}{2\ga^2}$.
 \ecl
\begin{proof}
 The proof of the claim follows directly  by substituting solving the equation
 for $\gr_1=\gr_2=1/2$.
\end{proof}
Finally we note that as $\SNR_i$
increases to infinity the NBS is always better than the NE.
\bcl If $SNR_1$
and $SNR_2$ are jointly increasing, while keeping the ratio  $\frac{SNR_1}{SNR_2}=z$ 
fixed. Then, there is  a constant $g$  such that for  $SNR_1>g$,
an FDM Nash Bargaining solution exists. \ecl
Proof: Define a function $h(x,z)$
\beq
\label{eq:def_f'}
h(x,z)=\min\left\{\gr:  \left(1+\frac{x}{\gr} \right)^\gr=1+z
\right\} .
\eeq
$z$ represents the constant ratio $x/y$.
The function $h(x,z)$ is monotonically decreasing to zero as a function of $x$ 
for any fixed value of $z$. 
Therefore, there is a constant $g$, such that for $x>g$ the inequality,  
$h\left(x,\frac{z}{\alpha }\right)+h\left(\frac{x}{z},\frac{1}{\beta z}\right)<1$ 
is satisfied. Since by definition of $f(x,y)$ we have $h(x,z)>f(x,y)$, 
the equation $f\left(x,\frac{x}{\alpha y }\right)+f\left(y,\frac{y}{\beta x}\right)<1$ 
also holds for all $x>g$ and $y=zx$.
\bcl If $SNR_1+SNR_2\leq \frac{1-\alpha-\beta}{\alpha\beta}$ there is no Nash bargaining solution.
\ecl
Proof: Nash Bargaining solution does not exists if
\beq
\left(1+\frac{SNR_1}{\rho}\right)^\rho \left(1+\frac{SNR_2}{1-\rho}\right)^{1-\rho}<
\left(1+\frac{SNR_1}{1+\alpha SNR_2}\right)\left(1+\frac{SNR_2}{1+\beta SNR_1}\right).
\eeq
\begin{proof}
The claim follows easily by applying the inequality 
$x^\rho y^{1-\rho}\leq \rho x+ \left(1-\rho\right)y$ on the left hand side of the above 
inequality and using the assumption.
\end{proof}
The following example
provides the intuition for the definitions of the game theoretic rate region, and 
the uniqueness  of the NBS using FDM strategies. It also clearly demonstrates the
relation between the competitive solution, the NBS and the game
theoretic rate region $\cR^G$. We have chosen $\SNR_1=20$ dB,
$\SNR_2=15$ dB, and $\ga=0.4, \gb=0.7$. Figure \ref{rate_region}
presents the FDM rate region, the Nash equilibrium point denoted by
$\*$, and a contour plot of $F(\gr)$. It can be seen that the
concavity of $NF(\gr)$ together with the convexity of the achievable
rate region implies that at there is a unique contour tangent to the
rate region. The tangent point is the Nash bargaining solution. We
can see that the NBS achieves rates that are 1.6 and 4 times higher
than the rates of the competitive Nash equilibrium rates for player
I and player II respectively. The game theoretic rate region is the
intersection of the information theoretic rate region with the
quadrant above the dotted lines.

\section{Bargaining over frequency selective channels under Mask constraint }
\label{freq_selective}
In this section we define a new cooperative game corresponding to the joint
FDM/TDM achievable rate region for the frequency selective $N$ users
interference channel. We limit
ourselves to the PSD mask constrained case since this case is actually
the more practical one. In real applications, the regulator
limits the PSD mask and not only the total power constraint. Let the
$K$ channel matrices at frequencies $k=1,...,K$ be given by
$\left<\mH_k:k=1,...,K\right>$. Each player is allowed to transmit
at maximum power $p\left(k\right)$ in the $k$'th frequency bin. In
non-cooperative scenario, under mask constraint, all players
transmit at the maximal power they can use. Thus, all players choose
the PSD, $\vp=\left<p_i(k): 1\le k \le K \right >$. The payoff for
user $i$ in the non-cooperative game is therefore given by:
\begin{equation} \label{eq_capacity}
 R_{iC}\left(\vp_1\right)=
 \sum_{k=1}^{K}\log_2\left(1+\frac{|h_i(k)|^2p_i(k)}{\sum_{j \neq i}
 |h_{ij}(k)|^2 p_j(k)+\gs_i^2(k)}\right).
 \end{equation}
 Here, $R_{iC}$ is the capacity available to player $i$ given a PSD mask constraint
 distributions $\vp$.  $\gs_i^2(k)>0$ is the noise presents at the $i$'th receiver at
 frequency $k$. Note that without loss of generality, and in order to simplify notation, 
we assume that the width of each bin is normalized to 1. 
 We know define the cooperative game $G_{TF}(N,K,\vp)$.
\begin{definition}
The FDM/TDM game $G_{TF}(N,K,\vp)$ is a game between $N$ players
transmitting over $K$ frequency bins under common PSD mask
constraint. Each user has full knowledge of the channel matrices
$\mH_k$. The following conditions hold: \begin{enumerate}
\item Player $i$ transmits using a PSD limited by $\left<p_i(k): \ k=1,...,K \right>$ satisfying $p_i(k)
\le p(k)$.
\item Strategies for player $i$ are vectors
$\vga=[\ga_{i1},...,\ga_{iK}]^T$ where $\ga_k$ is the proportion of
time the player uses the $k$'th frequency channel. This is the TDM
part of the strategy.
\item The utility of the $i$'th player is given by
\begin{equation} R_i=\sum_{k=1}^K R_i(k) = \sum_{k=1}^K \ga_{ik}\log_2
\left(1+\frac{|h_{ii}(k)|^2 p_i(k)}{\gs_i^2(k)}\right)
 \end{equation}
\end{enumerate}
 Note that interference is avoided by time sharing at each
frequency band, i.e only one player transmits at a given frequency
bin at any time. Furthermore, since at each time instance each
frequency is used by a single user, each user can transmit using
maximal power.
\end{definition}
The Nash bargaining  can be posed as an optimization problem
\begin{equation}
\begin{array}{c}
\textbf{max}\prod_{n=1}^N
\left(R_i(\boldsymbol{\alpha}_{i})-R_{iC}\right)\\
\textbf{subject to:} \bea{l}
            \sum_{i=1}^{N}\alpha_i(k)=1, \\
            \forall i,k \ \alpha_i(k)\geq0, \\
             \forall i \ R_{iC}\le R_i\left(\boldsymbol{\alpha}_{i}\right),
             \ena
\end{array}
\end{equation}
where,
\begin{equation}
 R_i\left(\boldsymbol{\alpha}_{i}\right)=
 \sum_{k=1}^{K}\alpha_i(k)\log_2\left(1+\frac{|h_i(k)|^2P_{max}\left(k\right)}{\gs_i^2(k)}\right)=
 \sum_{k=1}^{K}\alpha_i(k)R_{i}\left(k)\right).
\end{equation}
This problem is convex and therefore can be solved efficiently using
convex optimization techniques. To that end we explore the KKT
conditions for the problem. The Lagrangian of the problem
$f\left(\boldsymbol{\alpha}\right)$ is given by
\begin{equation}
  \begin{array}{c}
 f\left(\boldsymbol{\alpha}\right)=-\sum_{i=1}^N
\log\left(R_i(\boldsymbol{\alpha}_{i})-R_{iC}\right)
+\sum_{k=1}^{K}\lambda_{k}\left(\sum_{i=1}^{N}\alpha_i(k)-1\right) \\-\sum_{k=1}^{K}\sum_{i=1}^{N}\mu_i(k)\alpha_i(k)
-\sum_{i=1}^{N}\delta_i\left(\sum_{k=1}^K\alpha_i\left(k\right)R_i\left(k\right)-R_{iC}\right)
 \end{array}.
 \end{equation}
Taking the derivative with respect to the variable $\alpha_i(k)$ and comparing the result to zero, we get
 \begin{equation}
  \label{KKT1}
 \frac{R_{i}\left(k\right)}
     {R_i\left(\boldsymbol{\alpha}_{i}\right)-R_{iC}}=\lambda_{k}-\mu_i(k)-\delta_i
 \end{equation}
with the constraints
 \begin{equation}
 \label{KKT2}
 \sum_{i=1}^N\alpha_i\left(k\right)=1,
 \delta_i\left(R_i\left(\boldsymbol{\alpha}_{i}\right)-R_{iC}\right)\geq0,
 \mu_i(k)\alpha_i\left(k\right)=0, \lambda_{k}\geq0.
 \end{equation}
Based on ($\ref{KKT1}$, \ref{KKT2}) one can easily come to the following conclusions:
\begin{enumerate}
\item If there is a feasible solution then for all $i$, $\delta_i=0$.
\item Assume that a feasible solution exists. Then for all players sharing the frequency
bin $k$ ($\ga_i(k)>0$) we have $\mu_i(k)=0$, and
\begin{equation}
\label{KKT3}
 \frac{R_{i}\left(k\right)}
     {R_i\left(\boldsymbol{\alpha}_{i}\right)-R_{iC}}=\lambda_{k}, \forall k \texttt{ satisfying }\alpha_i\left(k\right)>0.
 \end{equation}
\item For all players that are not sharing the frequency bin $k$,($\alpha_i(k)=0$),
$\mu_i(k) \ge 0$. Therefore,
\begin{equation}
\label{KKT4}
 \frac{R_{i}\left(k\right)}
     {R_i\left(\boldsymbol{\alpha}_{i}\right)-R_{iC}} \le\lambda_{k}, \forall \texttt{k with }\alpha_i\left(k\right)=0.
 \end{equation}
\end{enumerate}
Clause (2) is very interesting. let $L_{ij}(k)=R_i(k)/R_j(k)$.
Assume that for users $i$,$j$ the values $L_{ij}(k)$
are all distinct. Then the two users can share at most a single frequency.
To see this note that in this case
\beq
 \frac{R_{i}\left(k\right)}
     {R_i\left(\boldsymbol{\alpha}_{i}\right)-R_{iC}}= \frac{R_{j}\left(k\right)}
     {R_j\left(\boldsymbol{\alpha}_{j}\right)-R_{jC}}
\eeq
and therefore
\beq
L_{ij}(k)=\frac{R_{i}\left(k\right)}{R_{j}\left(k\right)}=
\frac{R_i\left(\boldsymbol{\alpha}_{i}\right)-R_{iC}}{R_j\left(\boldsymbol{\alpha}_{j}\right)-R_{jC}}
\eeq
Since the right hand side is independent of the frequency $k$ and
$L_{ij}(k)$ are distinct, at most a single frequency can satisfy this
condition. This proves the following theorem: 
\bthm
\label{sharing_theorem} 
Assume that for all $i\neq j$ the values
$\left\{L_{ij}(k): k=1,...,K\right\}$ are all distinct. Then in the
optimal solution at most ${N} \choose {2}$ frequencies are shared
between different users. 
\ethm 
This theorem suggests, that when
${{N} \choose {2}}<<K$ the optimal FDM NBS is very close to the
joint FDM/TDM solution. It is obtained by allocating the common
frequencies to one of the users.

While general convex optimization techniques are useful for
computing the NBS, in the next section we will demonstrate that for
the two players case the solution can be computed much more
efficiently. Furthermore, we will show that in the optimal solution
only a single frequency is actually shared between the users even if
the $L_{ij}(k)$ are not distinct.

Finally we comment on the applicability of the method to the case where only 
fading statistics is known. In this case the coding strategy will change, and the achievable 
rate in the competitive case and the cooperative case are given by 
\beq
\bea{l}
{\tilde \mR_{iC}}\left(\vp_i\right)=
 \sum_{k=1}^{K}E\left[ \log_2\left(1+\frac{|h_i(k)|^2p_i(k)}{\sum_{j \neq i}
 |h_{ij}(k)|^2 p_j(k)+\gs_i^2(k)}\right)\right]\\
{\tilde R_i(\vga_i)}=\sum_{k=1}^K \ga_{ik}E\left[\log_2
\left(1+\frac{|h_{ii}(k)|^2 p_i(k)}{\gs_i^2(k)}\right)\right]
\ena
\eeq
respectively. All the rest of the discussion is unchanged, replacing $R_{iC}$ and $R_i(\vga_i)$
by ${\tilde \mR_{iC}},{\tilde R_i(\vga_i)}$ respectively.
\section{Computing the Nash bargaining solution for two players}
\label{two_players_section}
For the two players case the optimization
problem can be dramatically simplified. In this section we will
provide an $O(K \log_2 K)$ complexity algorithm (in the number of tones) for
computing the NBS optimal solution in a 2 users frequency selective
channel. Furthermore, we will show that the two players will share
at most a single frequency, no matter what the ratios between the users are.
To that end let,
$\alpha_1\left(k\right)=\alpha\left(k\right)$, and
$\alpha_2\left(k\right)=1-\alpha\left(k\right)$.
We also define the surplus of players I and II when using Nash bargaining solution as
$A=\sum_{m=1}^K\alpha\left(m\right)R_1\left(m\right)-R_{1C}$
 and $B=\sum_{,=1}^K\left(1-\alpha\left(m\right)\right)R_2\left(m\right)-R_{2C}$,
respectively. The ratio, $\Gamma=A/B$ is a threshold which is
independent of the frequency and is set by the optimal assignment. While $\Gamma$ is
a-priori unknown, it exists. Let $L(k)=R_1\left(k\right)/R_2\left(k\right)$.
Without loss of generality, assume that the rate
ratios $L(k), 1\leq k\leq K$ are sorted in decreasing order i.e.
$L(k)\geq L(k'), \forall k\leq k'.$ (This can be achieved by sorting the frequencies according
to $L(k)$.

We are now ready to define optimal assignment the
$\alpha$'s. Define three sets:
$S_1=\{m: L(m)>\Gamma, A>0, B>0\}, S_2=\{m: L(m)<\Gamma, A>0, B>0\}, 
S_c=\{m: L(m)=\Gamma, A>0, B>0\}$.
For all $m \in S_1$ $\ga(m)=1$. For all $m \in S_2$ $\ga(m)=0$. and for
$m  \in S_c$ $0 \le \ga(m) \le 1$.
Thus if the set $S_c$ is  empty, pure FDM is a Nash bargaining
solution.

Let $\Gamma_k$ be a moving
threshold defined by $\Gamma_k=A_k/B_k.$ where
\beq A_k=\sum_{m=1}^k
R_1\left(m\right)-R_{1C},   B_k=\sum_{m=k+1}^K
R_2\left(m\right)-R_{2C}.\eeq
$A_k$ is a monotonically increasing sequence, while $B_k$ is monotonically decreasing.
Hence, $\Gamma_k$ is also monotonically increasing. $A_k$ is the surplus of user I
respectively when frequencies $1,...,k$ are allocated to user I. Similarly $B_k$
is the surplus of user II when frequencies $k+1,...,K$
are allocated to user II.
Let
\beq
\bea{c}
k_{\min}=\min_k\left\{k: A_k \ge 0\right\};
k_{\max}=\min_k\left\{k:B_k < 0\right\}. \ena \eeq 
Since we are interested in feasible NBS, we must have positive surplus for both
users. Therefore, by the KKT equations, we obtain $k_{\min}\le k_{\max}$ and $
L(k_{\min}) \le \Gamma \le L(k_{\max})$. The sequence
$\{\Gamma_m: k_{min} \le m \le k_{max}-1\}$ is strictly
increasing, and always positive.  We first state two lemmas that are
essential for finding the optimal partition.
 \blm{} Assume that there is an NBS to the game. Then there is always a NBS satisfying
 that at most a  single bin $k_s$ is partitioned between the players, and
\beq \alpha(k)=\left\{
              \begin{array}{cc}
                1 & k<k_s  \\
                0 & k>k_s  \\
              \end{array}
              \right. .
            \eeq
 \elm{}
 \begin{proof}
By our assumption the sequence  $\{L(k):k=1,...,K\}$ is
monotonically decreasing (not necessarily strictly decreasing).
If there is a $k$ such that $L(k-1) < \Gamma < L(k)$ then the solution must be FDM type
by the KKT equations and we finish.
Otherwise assume that $L(k)=\Gamma$.
Since $\Gamma_k$ is strictly increasing and $L(k)$ is non-increasing
there is at most a unique $k$ such that $\Gamma_{k-1} \le L(k)=\Gamma <\Gamma_k$.
If no such $k$ exists then the users can only share $k_{\max}$ since for all $k \le k_{\max}$
\[
\frac{A_k}{B_k} \le \Gamma
\]
and the only way to get something allocated to user II is by sharing $k_{\max}$.
Otherwise such a $k \le k_{\max}$ exists.
By definition of $\Gamma_k$ we have
\[
\frac{A_{k-1}}{B_{k-1}} \le L(k) < \frac{A_{k}}{B_{k}}.
\]
Simple substitution yields
\[
\frac{A_{k-1}}{B_{k-1}} \le L(k) <
\frac{A_{k-1}+R_1(k)}{B_{k-1}-R_2(k)}=\frac{A_k}{B_k}.
\]
Since $k_{\min}\le k < k_{\max}$ the denominator on the RHS is
positive. Since for $a,b,c,d>0$ the function $\frac{a+xb}{c-xd}$ is
increasing with $0 \le x$ as long as the denominator is positive, we
obtain that by continuity there is a unique $\gz$ such that
\[
L(k)=\frac{A_{k-1}+\gz R_1(k)}{B_{k-1}-\gz R_2(k)}.
\]
But $B_{k-1}-\gz R_2(k)=B_k+(1-\gz)R_2(k)$
so that $\gz$ satisfies
\[
\Gamma=L(k)=\frac{A_{k-1}+\gz R_1(k)}{B_k+(1-\gz)R_2(k)}.
\]
Setting $\ga(m)=1$ for $m<k, \ga(k)=\gz$ and $\ga(m)=0$ for $m>k$ we
obtain a solution of the KKT equations. Note that when there are
multiple values of $k$ such that $L(k)=\Gamma$, we only showed that
there is an NBS solution where a single frequency is shared.
\end{proof}
While the threshold $\Gamma$ is unknown, one can use the sequences
$\Gamma_k$ and $L(k)$.

\begin{figure}
  \begin{center}
    \mbox{\psfig{figure=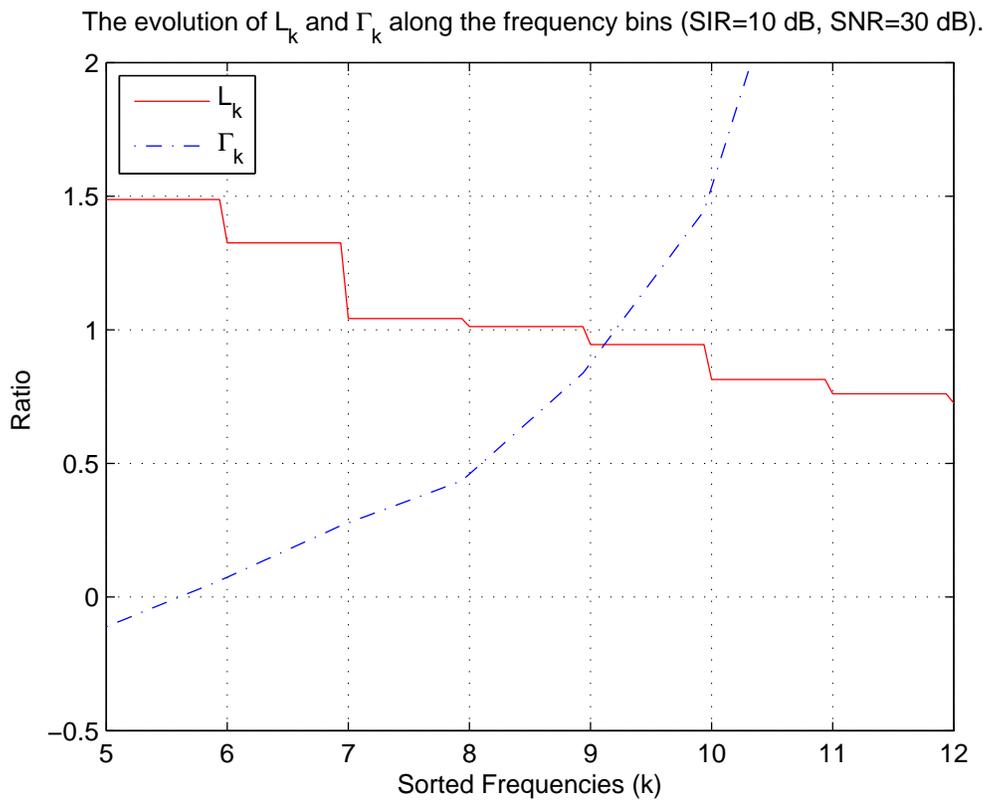,width=0.8\textwidth}}
 \end{center}
   \caption{Sorted $L(k)$ and $\Gamma_k$.}
   \label{threshold}
 \end{figure}

If there is a Nash bargaining solution, let $k_s$ be the frequency bin that
is shared by the players. Then, $k_{\min}\leq k_s\leq k_{\max}$.
Since, both players must have a positive gain in the game 
($A>A_{k_{\min}-1}$,$B>B_{k_{\max}}$). Let $k_s$ be the
smallest integer such that $L(k_s)<\Gamma_{k_s}$, if such $k_s$ exists. Otherwise let 
$k_s=k_{\max}$.
 \blm{} The following two statements provide the solution
\begin{description}
\item [1] If a Nash bargaining solution exists for $k_{\min}\leq k_s < k_{\max}$, 
then $\alpha\left(k_s\right)$ is given by
$\alpha\left(k_s\right)=max\{0, g\},$ where \beq g=
1+\frac{B_{k_s}}{2 R_2\left(k_s\right)}\left(1-
\frac{\Gamma_{k_s}}{L(k_s)}\right). \eeq
\item [2] If a Nash bargaining solution exists and there is no such $k_s$, then
$k_s=k_{max}$ and  $\alpha\left(k_s\right)=g$.
\end{description}
\elm
\begin{proof}
To prove 1 note that since $\Gamma_{k_s-1}\leq L(k_s)\leq\Gamma_{k_s}$,
$\alpha\left(k_s\right)$ is the solution to the equation
$L(k_s)=\frac{A_{k_s}-\left(1-\alpha\left(k_s\right)\right)R_1\left(k_s\right)}{B_{k_s}+\left(1-\alpha\right)R_2\left(k_s\right)}$.
By simple mathematical manipulation,  we get
$\alpha\left(k_s\right)=g.$ Since, $L(k)\leq \Gamma_{k_s}$, $g\leq1$.
If $g$ is negative, we set $\alpha\left(k_s\right)=0$, since $k_s$ is the
smallest integer such that $L(k_s)<\Gamma_{k_s}$. Note, that in this case 
the Nash bargaining solution is given by pure FDM strategies.

To prove 2 note that since $k_s=k_{\max}$ and $\Gamma_k$ is increasing for 
$k_{\min}\leq k < k_{\max}$, we must have that 
$\Gamma_{k_{\max}-1}\leq\Gamma=L(k_{\max})$. Therefore, the
only possibility that there is a solution is if  $k_s=k_{max}$, and
$\alpha\left(k_s\right)=g\geq 0$.
\end{proof}
Based on the pervious lemmas the algorithm is described in table
\ref{two_players_table}. In the first stage the algorithm computes $L(k)$
and sorts them in a non increasing order. Then
$k_{\min}, k_{\max}, A_k,$ and $B_k$ are computed. In the second stage the
algorithm computes  $k_s$ and $\vga$. 
Figure \ref{threshold} demonstrates the situation when $SNR=30$dB and SIR is 
$10$dB. In this case $k_{\max}=10$ since $B_{11}$ becomes negative. 
Also $\Gamma_8<L(9)<\Gamma_9$. Therefore, 
only frequency $9$ might be shared between the users.
The algorithm computes a Nash
bargaining solution if it exists, even in the case that $L(k)$ is
not a strictly decreasing sequence. However, reordering the bins with
identical ratio may provides a different solution, with the same
capacity gain for each player.

\section{Simulations}
In this section we compare in simulations the Bargaining solution to the
competitive solution for various situations with medium interference.
The simulations are done both for flat slow fading and for frequency selective fading.
First, we demonstrate the effect of the channel matrix and the signal to noise ratio on
the gain of the NBS for flat fading channel. Then we performed extensive simulations
that demonstrate the advantage of the NBS over the competitive approach for the frequency
selective fading channel, as a function of the mean interference power.
\subsection{Flat fading}
\begin{figure}
  \begin{center}
    \mbox{\psfig{figure=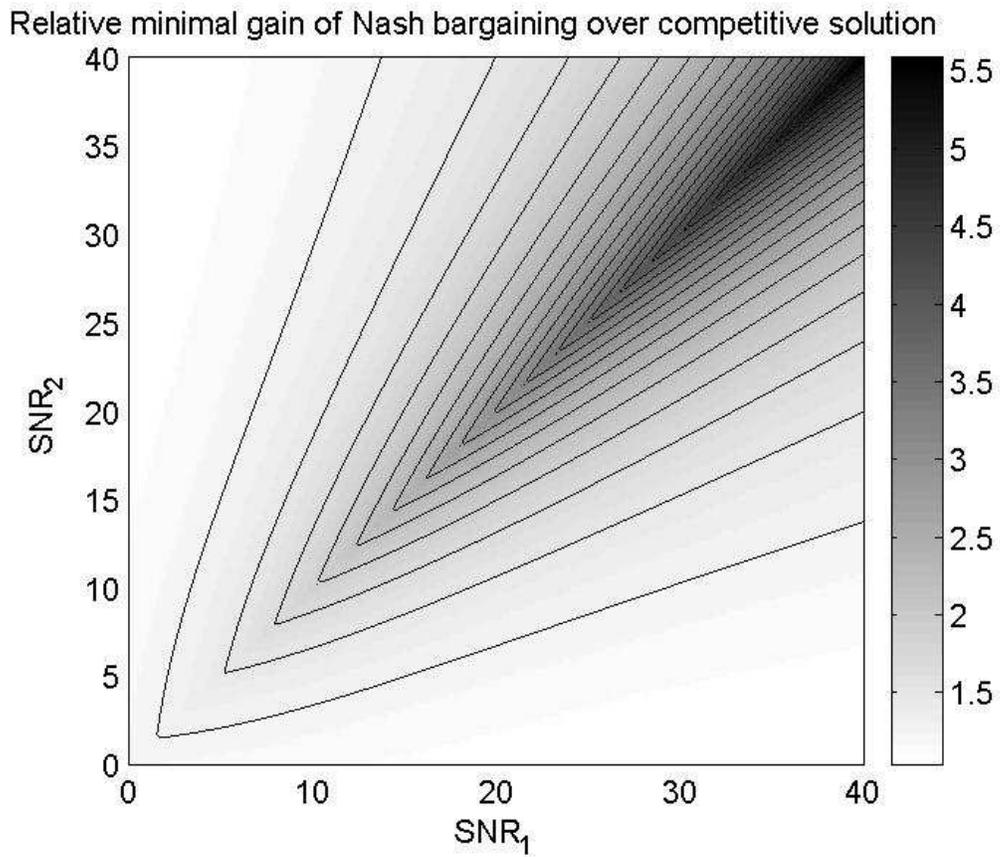,width=0.8\textwidth}}
 \end{center}
   \caption{Per user price of anarchy (relative improvement of NBS sum rate over NE), as a function of SNR. $\ga=\gb=0.7$.}
   \label{snra7b7}
 \end{figure}

\begin{figure}
 \begin{center}
    \mbox{\psfig{figure=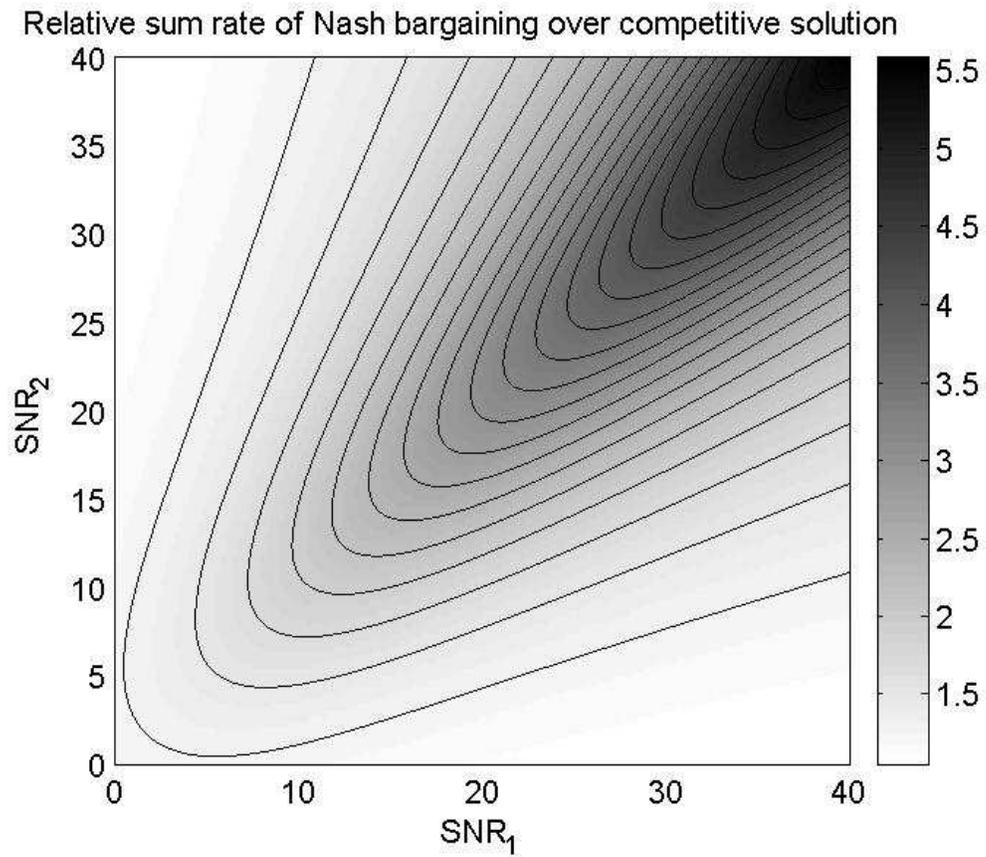,width=0.8\textwidth}}
 \end{center}
    \caption{Price of anarchy, as a function of SNR. $\ga=\gb=0.7$.}
 \label{sum_rate_SNRa7b7}
 \end{figure}

\begin{figure}
  \begin{center}
    \mbox{\psfig{figure=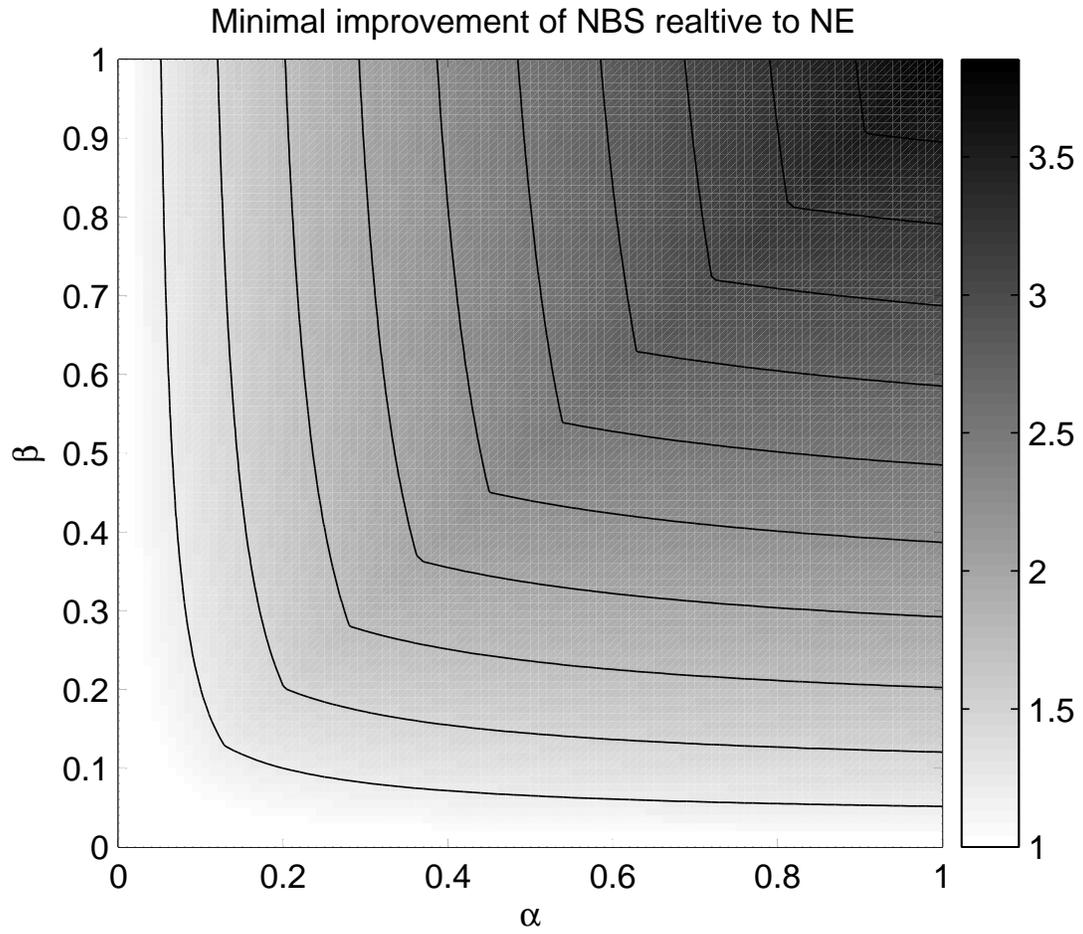,width=0.8\textwidth}}
 \end{center}
  \caption{Per user price of anarchy. SNR=20 dB.}
  \label{snr_minCH20}
 \end{figure}

\begin{figure}
  \begin{center}
    \mbox{\psfig{figure=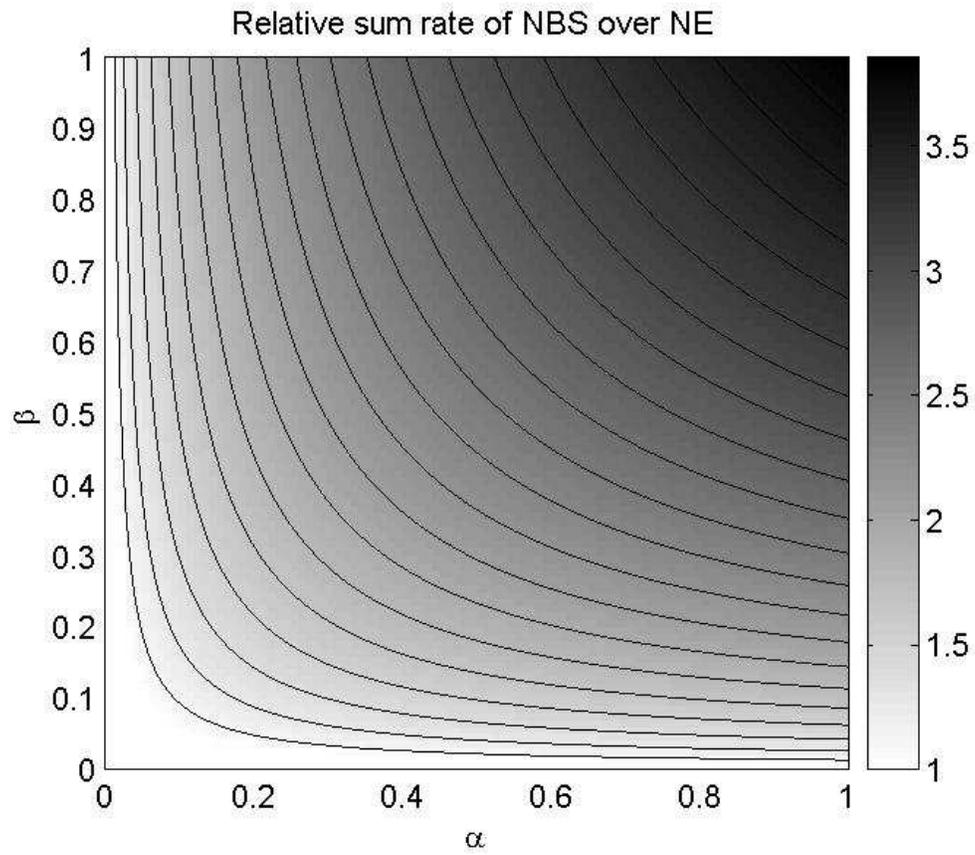,width=0.8\textwidth}} \end{center}
    \caption{Sum rate price of anarchy as a function of interference power. SNR=20 dB.}
\label{sum_rate_snrCH20}
 \end{figure}

We have  tested the gain of the Nash bargaining solution relative to the Nash
 equilibrium competitive rate pair as a function of channel coefficients as
 well as signal to noise ratio for the flat fading channel.
 To that end we define the minimum relative
improvement describing the individual price of anarchy by:
 \beq
 \label{min_delta}
 \gD_{\min}=\min\left\{
 \frac{R^{NBS}_1}{R^c_1},\frac{R^{NBS}_2}{R^c_2}
   \right\}
 \eeq
 and the usual price of anarchy \cite{papadimitriou2001}, describing total loss due to lack of cooperation by
 \beq
 \label{sum_delta}
 \gD_{sum}=\frac{R^{NBS}_1+R^{NBS}_2}{R^c_1+R^c_2}.
 \eeq
 In the first set of experiments we have fixed $\ga,\gb$ and
 varied $\SNR_1,\SNR_2$ from 0 to 40 dB in steps of 0.25dB.
 Figure \ref{snra7b7} presents $\gD_{\min}$ for an
 interference channel with $\ga=\gb=0.7$. We can see that for high
 SNR we obtain significant improvement. Figure \ref{sum_rate_SNRa7b7}
 presents the relative sum rate improvement $\gD_{sum}$ for the same channel.
 We can see that the achieved rates are 5.5 times those of the competitive
 solution.
 We have now studied the effect of the interference coefficients on the Nash
 Bargaining solution. We have set the signal to additive white Gaussian noise
 ratio for both users to 20 dB, and varied $\ga$ and $\gb$ between $0$ and $1$.
 Similarly to the previous case we present the minimal price of anarchy per user
$\gD_{\min}$ and the sum rate price of anarchy $\gD_{sum}$. The results are shown
in figures \ref{snr_minCH20},\ref{sum_rate_snrCH20}. We can clearly see that even with SINR
of $10$ dB we obtain 50 percent capacity gain per user.

\subsection{Frequency selective Gaussian channel}
\begin{figure}
  \begin{center}
    \mbox{\psfig{figure=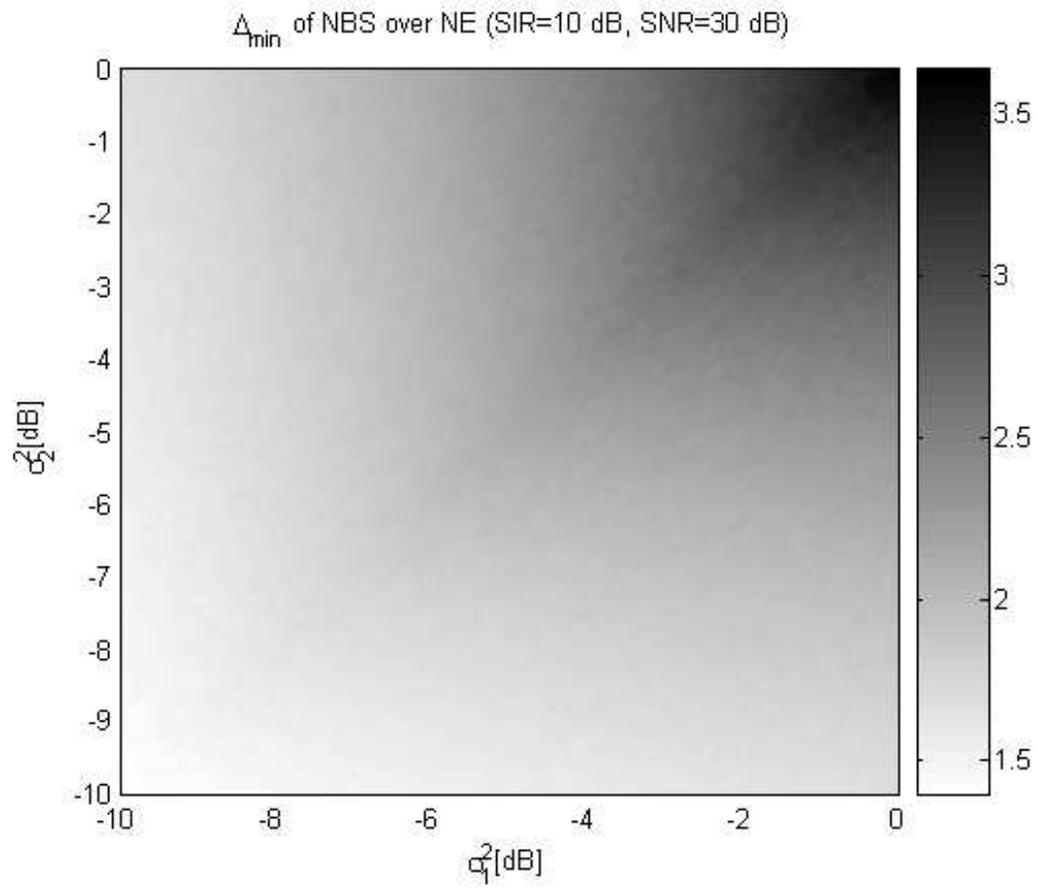,width=0.8\textwidth}}
 \end{center}
  \caption{Per user price of anarchy for frequency selective Rayleigh fading channel.
  SNR=30 dB.}
  \label{NBS}
 \end{figure}

In this experiment We demonstrate the advantage of the Nash bargaining
solution over competitive approaches for a frequency selective interference channel.
We assumed that two users having direct channels that are standard Rayleigh fading channels
($\gs^2=1$), with SNR=30 dB, suffer from interference, with SINR of each
user into the other channel ($h_{ij}$) was varied from 10 dB to 0 dB
($\gs_{h_{ij}}=0.1,...1$).
We have used 32 frequency bins. At each pair of variances $\gs_1^2=\gs_{h_{21}}^2,\gs_2^2=\gs_{h_{12}}^2$
we randomly picked 25 channels (each comprising of 32 2x2 matrices).
The results of the minimal relative improvement (\ref{min_delta})
are depicted in figure \ref{NBS}.
We can clearly see that the relative gain of the Nash bargaining solution over the
competitive solution is 1.5 to 3.5 times, which clearly demonstrates the merrits of
the method.

 \section{Conclusions}
In this paper we have defined the tic rate region for the
interference channel. The region is a subset of the rate region of the
interference channel. We have shown that a specific point in the rate region
given by the Nash bargaining solution is better than other points in the context of bargaining theory. We have shown conditions for the existence of such a
point in the case of the FDM rate region.  We have shown that computing the Nash bargaining solution
over a frequency selective channel can be described as a convex optimization problem.
Moreover, we have provided a very simple algorithm for solving the problem in the 2xK case
that is $O(K \log_2 K)$, where $K$ is the number of tones. Finally, we have demonstrated through
simulations the significant improvement of the cooperative solution over the
competitive Nash equilibrium.

The adaptation of game theory approach for rate allocation in existing wireless and
wireline system is very appealing. In many wireless LAN systems there is a central access point
with full knowledge on the channel transfer functions. Moreover, it has been recognized by the
802.11 committee that radio resource management is importnat, especially when multiple
networks are interfering with other. Knowledge of the transfer functions allows the access
point to allocate the band for the subscribers on the uplink. Moreover,
the results here can be extended to MIMO systems as well as for networks with
multiple access points.

\bibliographystyle{ieeetr}
\bibliography{cioffi_bib,interference,DSL_bib,dsm}

\newcommand{\noopsort}[1]{} \newcommand{\printfirst}[2]{#1}
  \newcommand{\singleletter}[1]{#1} \newcommand{\switchargs}[2]{#2#1}
\begin{thebibliography}{10}

\bibitem{leshem2006}
A.~Leshem and E.~Zehavi, ``Bargaining over the interference channel,'' in {\em
  Proc. IEEE ISIT}, pp.~2225--2229.

\bibitem{cover}
T.M. Cover and J.~A. Thomas, {\em Elements of Information Theory}.
\newblock New York, NY: John Wiley and Sons, 1991.

\bibitem{Meulen94}
E.C. {van der Meulen}, ``Some reflections on the interference channel,'' in
  {\em Communications and Cryptography: Two Sides of One Tapestry} (R.E.
  Blahut, D.~J. Costell, and T.~Mittelholzer, eds.), pp.~409--421, Kluwer,
  1994.

\bibitem{ahlswede73}
R.~Ahlswede, ``Multi-way communication channels,'' in {\em Proceedings of 2nd
  International Symposium on Information Theory}, pp.~23--52, Sept. 1973.

\bibitem{cheng93}
R.S. Cheng and S.~Verdu, ``On limiting characterizations of memoryless
  multiuser capacity regions,'' {\em IEEE Trans. on Information Theory},
  vol.~39, pp.~609--612, Mar. 1993.

\bibitem{han81}
T.S. Han and K.~Kobayashi, ``A new achievable rate region for the interference
  channel,'' {\em IEEE Trans. on Information Theory}, vol.~27, pp.~49--60, Jan.
  1981.

\bibitem{sason2004}
I.~Sason, ``On achievable rate regions for the {G}aussian interference
  channel,'' {\em IEEE Trans. on Information Theory}, vol.~50, pp.~1345--1356,
  June 2004.

\bibitem{sato81}
H.~Sato, ``The capacity of the {G}aussian interference channel under strong
  interference,'' {\em IEEE Trans. on Information Theory}, vol.~27,
  pp.~786--788, nov 1981.

\bibitem{yu04}
W.~Yu, W.~Rhee, S.~Boyd, and J.M. Cioffi, ``Iterative waterfilling for
  {G}aussian vector multiple-access channels,'' {\em IEEE Transactions on
  Information Theory}, vol.~50, no.~1, pp.~145--152, 2004.

\bibitem{laufer2005}
A.~Laufer and A.~Leshem, ``Distributed coordination of spectrum and the
  prisoner's dilemma,'' in {\em Proc. of the First IEEE International Symposium
  on New Frontiers in Dynamic Spectrum Access Networks - DySPAN 2005}, pp.~94
  -- 100, 2005.

\bibitem{laufer2005a}
A.~Laufer, A.~Leshem, and H.~Messer, ``Game theoretic aspects of distributed
  spectral coordination with application to {DSL} networks.'' arXiv:cs/0602014,
  2005.

\bibitem{etkin2005}
R.~Etkin, A.~Parekh, and D.~Tse, ``Spectrum sharing for unlicensed bands,'' in
  {\em Proc. of the First IEEE International Symposium on New Frontiers in
  Dynamic Spectrum Access Networks - DySPAN 2005}, pp.~251 -- 258, 2005.

\bibitem{owen}
G.~Owen, {\em Game theory}.
\newblock Academic Press, third~ed., 1995.

\bibitem{mathur2006}
S.~Mathur, L.~Sankaranarayanan, and N.B. Mandayam, ``Coalitional games in
  gaussian interference channels,'' in {\em Proc. IEEE ISIT}, pp.~2210 -- 2214,
  2006.

\bibitem{han2005}
Z.~Han, Z.~Ji, and K.J.R. Liu, ``Fair multiuser channel allocation for {OFDMA}
  networks using the {N}ash bargaining solutions and coalitions,'' {\em IEEE
  Trans. on Communications}, vol.~53, pp.~1366--1376, Aug. 2005.

\bibitem{yu2002}
W.~Yu, G.~Ginis, and J.M. Cioffi, ``Distributed multiuser power control for
  digital subscriber lines,'' {\em IEEE Journal on Selected areas in
  Communications}, vol.~20, pp.~1105--1115, june 2002.

\bibitem{yu2000}
Wei Yu and J.M. Cioffi, ``Competitive equilibrium in the {G}aussian
  interference channel",,'' in {\em Proc. of ISIT}, p.~431, June 2000.

\bibitem{chung2002}
S.T. Chung, J.~Lee, S.J. Kim, and J.M. Cioffi, ``On the convergence of
  iterative waterfilling in the frequency selective {G}aussian interference
  channel,'' {\em Preprint}, 2002.

\bibitem{luo2005}
Z.-Q. Luo and J.-S. Pang, ``Analysis of iterative waterfilling algorithm for
  multiuser power control in digital subscriber lines,'' {\em EURASIP Journal
  on Applied Signal Processing on Advanced Signal Processing Techniques for
  Digital Subscriber Lines}.

\bibitem{papadimitriou2001}
C.~Papadimitriou, ``Algorithms, games and the internet,'' in {\em Proc. of
  34'th ACM symposium on theory of computing}, pp.~749--753, 2001.

\end{thebibliography}
\newpage

\begin{table}
\centering
\caption{Algorithm for computing the 2x2 frequency selective NBS}
\begin{tabular}{||l||}
\hline
\hline
{\bf: Initialization:}  Sort the ratios $L(k)$
in decreasing order. \\
Calculate the values of $A_k,B_k$ and $\Gamma_k, k_{\min}, k_{\max}$, \\
\hline
If $k_{\min}>k_{\max}$ no NBS exists. Use competitive solution. \\
Else \\
\quad For $k=k_{\min}$ to $k_{max}-1$ \\
 \quad \quad if $L(k)\leq\Gamma_k$. \\
\quad \quad \quad Set $k_s=k$ and  $\alpha'$s according to the
lemmas-This is NBS. Stop\\
\quad \quad End \\
\quad End \\
\quad If no such $k$ exists, set $k_s=k_{\max}$ and calculate $g$. \\
\quad If $g\geq0$ set $\alpha_{k_s}=g, \alpha(k)=1$, for $k<k_{\max}$. Stop. \\
\quad Else ($g<0$) \\
 \quad \quad There is no NBS. Use competitive solution. \\
 \quad End. \\
End \\
\hline
\hline
\end{tabular}
\label{two_players_table}
\end{table}


\end{document}